%% file: apssamp.tex
\documentclass[%
 reprint,
 amsmath,amssymb,
 aps,
]{revtex4-2}

\usepackage{graphicx}
\usepackage{dcolumn}
\usepackage{bm}
\DeclareMathOperator\arctanh{arctanh}

\usepackage[per-mode=symbol]{siunitx}
\usepackage{braket}
\usepackage{relsize}

\begin{document}

\preprint{APS/123-QED}

\title{Observation of the Decrease of Larmor Tunneling Times with Lower Incident Energy}

\author{David C. Spierings}
\email{dspierin@physics.utoronto.ca}
\author{Aephraim M. Steinberg}%
 \altaffiliation[Also at ]{Canadian Institute For Advanced Research, Toronto, Ontario, Canada.}
\affiliation{%
 Centre for Quantum Information and Quantum Control, Department of Physics, University of Toronto, 60 St.\ George Street, Toronto, Ontario M5S 1A7, Canada
}%


\begin{abstract}
How much time does a tunneling particle spend in a barrier? A Larmor clock, one proposal to answer this question, measures the interaction between the particle and the barrier region using an auxiliary degree of freedom of the particle to clock the dwell time inside the barrier. We report on precise Larmor time measurements of ultra-cold $^{87}$Rb atoms tunneling through an optical barrier, which confirm longstanding predictions of tunneling times. We observe that atoms generally spend less time tunneling through higher barriers and that this time decreases for lower energy particles. For the lowest measured incident energy, at least \SI{90}{\percent} of transmitted atoms tunneled through the barrier, spending an average of \SI{0.59\pm0.02}{\ms} inside.  This is \SI{0.11\pm0.03}{\ms} faster than atoms traversing the same barrier with energy close to the barrier's peak and \SI{0.21\pm0.03}{\ms} faster than when the atoms traverse a barrier with \SI{23}{\percent} less energy.

\end{abstract}

\maketitle


Tunneling is one of the most famous quantum phenomena and plays a central role in many physical contexts.  Despite its ubiquity, certain aspects of tunneling remain enigmatic. 
How much time does tunneling take? This has been a provocative question for decades, not only because no definitive answer has emerged, but also because there is no consensus on a definition for this time \cite{hauge1989review,landauer1994review,steinberg1997review}. Classical intuition would lead one to believe that numerous measures of time should probe the same quantity.  Yet, the lack of definite trajectories in quantum mechanics causes seemingly equivalent approaches to yield strikingly different results. The continued work on tunneling times is thus not motivated by the search for a unique timescale, but rather by the hope that a few useful and physically significant definitions can be distilled from the countless proposals, and their relationships made clear. We emphasize in particular the distinction between two categories of tunneling time definitions: `arrival' and `interaction' times.

Arrival times seek to determine the moment at which a transmitted particle emerges on the far side of a barrier. 
Early work on the tunneling time problem focused on calculating the group delay, which tracks a wavepacket's peak and can be superluminal or even negative \cite{ranfagni1991,enders1992,steinberg1993,spielmann1994}
without violating relativistic causality \cite{brillouin,steinberg1997review}. More recently, the `attoclock' has been demonstrated as a means of measuring the arrival time of an electron escaping from a bound state of the atom's Coulomb potential \cite{eckle2008attosecond}. There, the rotating polarization of an ionizing pulse provides a kick to the freed electron, thereby encoding the exit time of the tunneling electron in its final momentum.  By comparing the exit time to the instant the electric field reaches its maximum, when the ionization probability also reaches its peak, a tunneling delay can be extracted.  While a recent measurement \cite{Sainadh2019,sainadh2020attoclock} demonstrates no appreciable delay in the tunneling process, other experiments \cite{landsman2014,Camus2017,fortun2016direct} report non-zero delays that remain unexplained \cite{Torlina2015,ni2016tunneling,Zimmermann2016,hofmann2019attoclock}.

Interaction times strive instead to describe how much time a particle \textit{spends} inside the barrier region. Here we consider the Larmor clock \cite{baz1966lifetime,Rybachenko1966}.
Early measurements of Larmor times for tunneling particles include work in analog optical systems \cite{deutsch1996optical,balcou1997dual} and with neutrons \cite{hino1999measurement}.
Recently, we made a Larmor time measurement \cite{Ramos2020} with ultra-cold atoms that disentangled the nonzero time spent inside the barrier from the back-action of the measurement. These results constituted an important milestone, but were severely limited by systematic errors as large as \SI{40}{\percent} of the measured times as well as uncertain effective temperatures of the incident wavepacket, \SI{2\pm1}{\nano\kelvin}, which meant that the portion of transmission due to tunneling may have only been a small minority.

Here, we report new Larmor time measurements with systematic errors reduced to typically \SI{4}{\percent} of the measured values and with lower-temperature, stable wavepacket preparation, \SI{1.3\pm0.2}{\nano\kelvin}. In addition, thanks to more sensitive imaging, capable of reliably counting tens of atoms (another order of magnitude improvement), we can now observe transmission at energies such that tunneling is the dominant means of passage through the barrier. We observe that the time particles spend interacting with the barrier decreases with their energy below it, while the back-action grows. This is observable only because a large majority of atoms tunnel through the barrier for the lowest measured energies.  Furthermore, we demonstrate that particles spend less time tunneling through a higher barrier.


A Larmor clock uses an auxiliary degree of freedom of the tunneling particle to measure the dwell time inside the barrier.  The dwell time is simply the probability of finding a particle in a region integrated over all times, which in one dimension is given by
\begin{align}
\label{eq:dwell}
\tau_d&=\int^\infty_{-\infty}dt\int^{y_2}_{y_1}dy|\psi(y,t)|^2\equiv\int^\infty_{-\infty}dt\bra{\psi}\Theta_d\ket{\psi},
\end{align}
where $y_1$ and $y_2$ are the boundaries of the region of interest, $\Theta_d\equiv\int^{y_2}_{y_1}dy\ket{y}\bra{y}$ is a projector onto the barrier region, and $\psi(y,t)$ is the wavefunction.  The dwell time is a clear-cut definition of the time spent in a region for any probabilistic evolution, but in the case of tunneling does not distinguish between reflected and transmitted subensembles. A Larmor clock implements a measurement of time conditioned on transmission or reflection by having the measuring device carried by each particle and arranging the measurement interaction so that it does not significantly affect the motion \cite{Ae1995_time}.

\begin{figure}[t]
\includegraphics[width=\columnwidth]{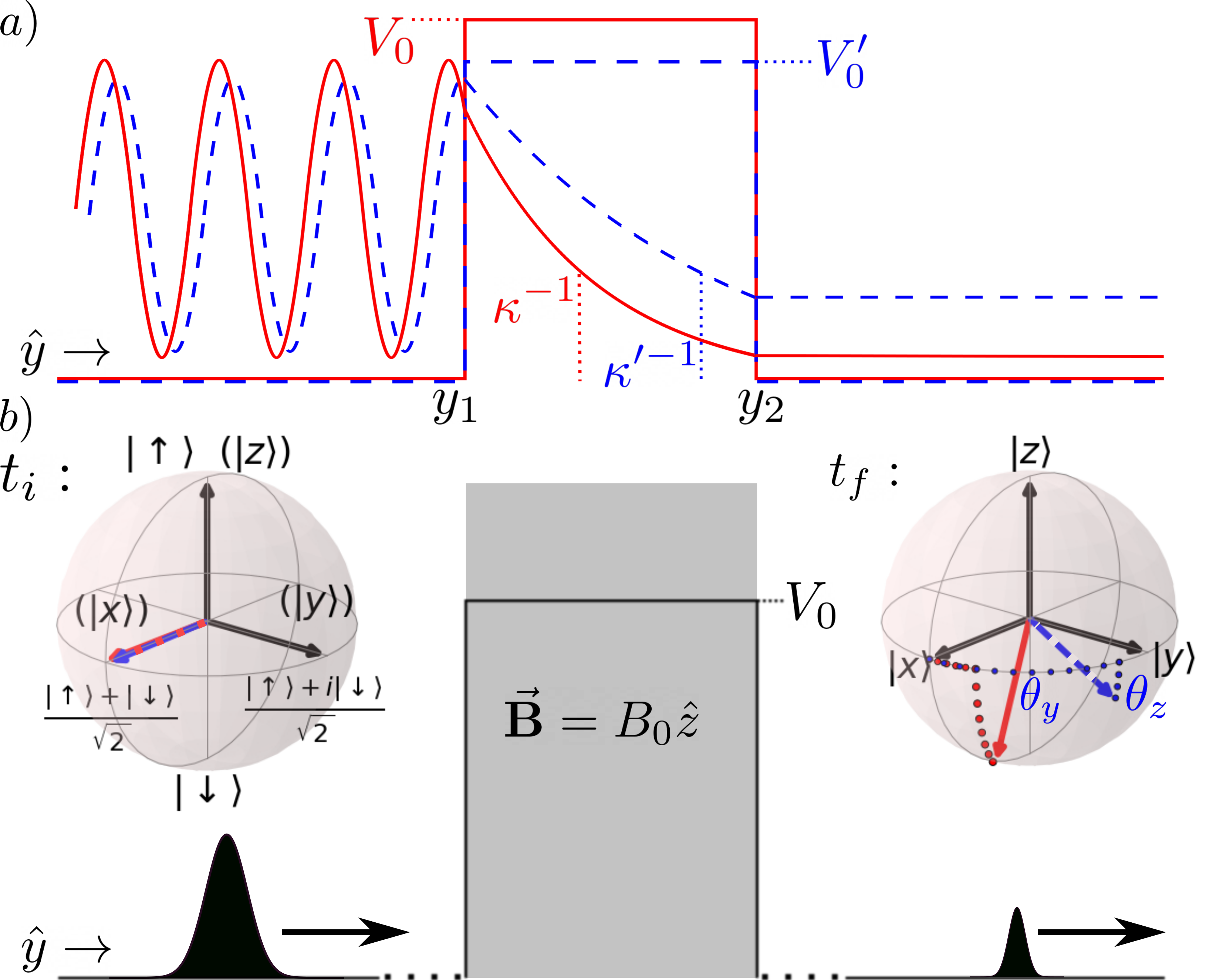}
\caption{\label{fig:larmorthought}a) Illustrations of $|\psi(y)|^2$ for a wave of energy $E$ incident on square barriers of two heights. The dwell time in the red (solid) scenario, with a higher barrier, is lower than in the blue (dashed) case, with a lower barrier. 
b) The standard Larmor experiment starts with a spin-1/2 particle (here an electron), polarized along the x-axis, incident on a square barrier.  A magnetic field, localized to the barrier region and pointing along the z-axis, causes the spin to precess while in the barrier.  The angles of precession, $\theta_y$, and alignment, $\theta_z$, of the transmitted wavepacket determine the Larmor tunneling times when the energy of the wavepacket is below the barrier. The right Bloch sphere illustrates the rotations resulting from a Larmor measurement of the scenarios shown in (a).}
\end{figure}

The paradigmatic Larmor experiment considers an ensemble of spin-1/2 particles incident on a barrier as illustrated in Fig. \ref{fig:larmorthought}b, drawn for an electron.  Before interacting with the barrier, each particle's spin is polarized along the x-axis of the Bloch sphere.  A magnetic field, pointing in the z-direction and localized only within the barrier, causes the spin of the particle to precess while inside the barrier.  By measuring the precession angle $\theta_y$ after the scattering event for either reflected or transmitted particles, the dwell time conditioned on these final states can be extracted from $\tau_y=\theta_y/\omega_L$, where $\omega_L$ is the Larmor frequency.  In addition to the `in-plane' angle, $\theta_y$, an `out-of-plane' angle, $\theta_z$, results from the preferential transmission of the spin component anti-parallel to the magnetic field, which experiences an effectively lower barrier due to the magnetic potential energy \cite{buttiker1983larmor}.  Accordingly, a second time $\tau_z=\theta_z/\omega_L$ can be defined. 

The meaning of these two times becomes clear by viewing the Larmor clock as a von Neumann measurement \cite{Ae1995_cond,Ae1995_time}.  The interaction Hamiltonian of the Larmor clock, $H_{int}\sim\boldsymbol{S}\cdot\boldsymbol{B}\sim S_z B_0\Theta_{d}$, couples the spin of each particle to the projector onto the barrier region.  In this way, the spin acts as the pointer of a measurement apparatus observing whether the system (each particle) is inside the barrier. While $S_z$, the occupation difference between $\ket{\uparrow}$ and $\ket{\downarrow}$, represents the pointer momentum (i.e. the generator of pointer translations), the conjugate pointer position is the phase difference between $\ket{\uparrow}$ and $\ket{\downarrow}$.  Accordingly, we associate $\tau_y$ with the time spent inside the barrier and $\tau_z$ with the measurement back-action due to the effect of this interaction on the tunneling probability.  In the limit of a weak measurement, $B_0\rightarrow0$,  the Larmor times for transmission can be written \cite{buttiker1983larmor}
\begin{align}
\label{eq:tauy}
\tau_y+i\tau_z&=-\hbar\frac{\mathlarger{\partial}\phi}{\mathlarger{\partial} V_0}+i\hbar\frac{\mathlarger{\partial} \ln(|t|)}{\mathlarger{\partial} V_0}=i\hbar\frac{\mathlarger{\partial} \ln t}{\mathlarger{\partial} V_0},
\end{align}
where $\phi$ is the phase of the transmission amplitude $t$, and $V_0$ is the energy of the barrier.  It turns out that the Larmor times are equivalent to the weak value of the projector $\Theta_d$ \cite{AAV,Ae1995_time}.

Figure \ref{fig:larmorthought}a illustrates that a particle with energy $E$ and penetration depth $\kappa^{-1}\propto(V_0-E)^{-1/2}$ incident on a barrier of height $V_0$ has a lower probability to be in the forbidden region than the same particle would when incident on a lower energy barrier, $V_0'$. Hence, the dwell time in the case with the higher barrier is shorter and similarly the dwell time for a fixed barrier height is shorter for particles with less energy. Here we demonstrate that these trends hold for the Larmor tunneling time $\tau_y$, a fact predicted nearly $40$ years ago \cite{buttiker1983larmor,Falck1988}. One might find it reasonable that lower energy particles penetrate less deeply and therefore are {\it reflected} without spending much time in the barrier. No such argument applies to transmitted particles which traverse the entire barrier.

\begin{figure}[t]
\includegraphics[width=\columnwidth]{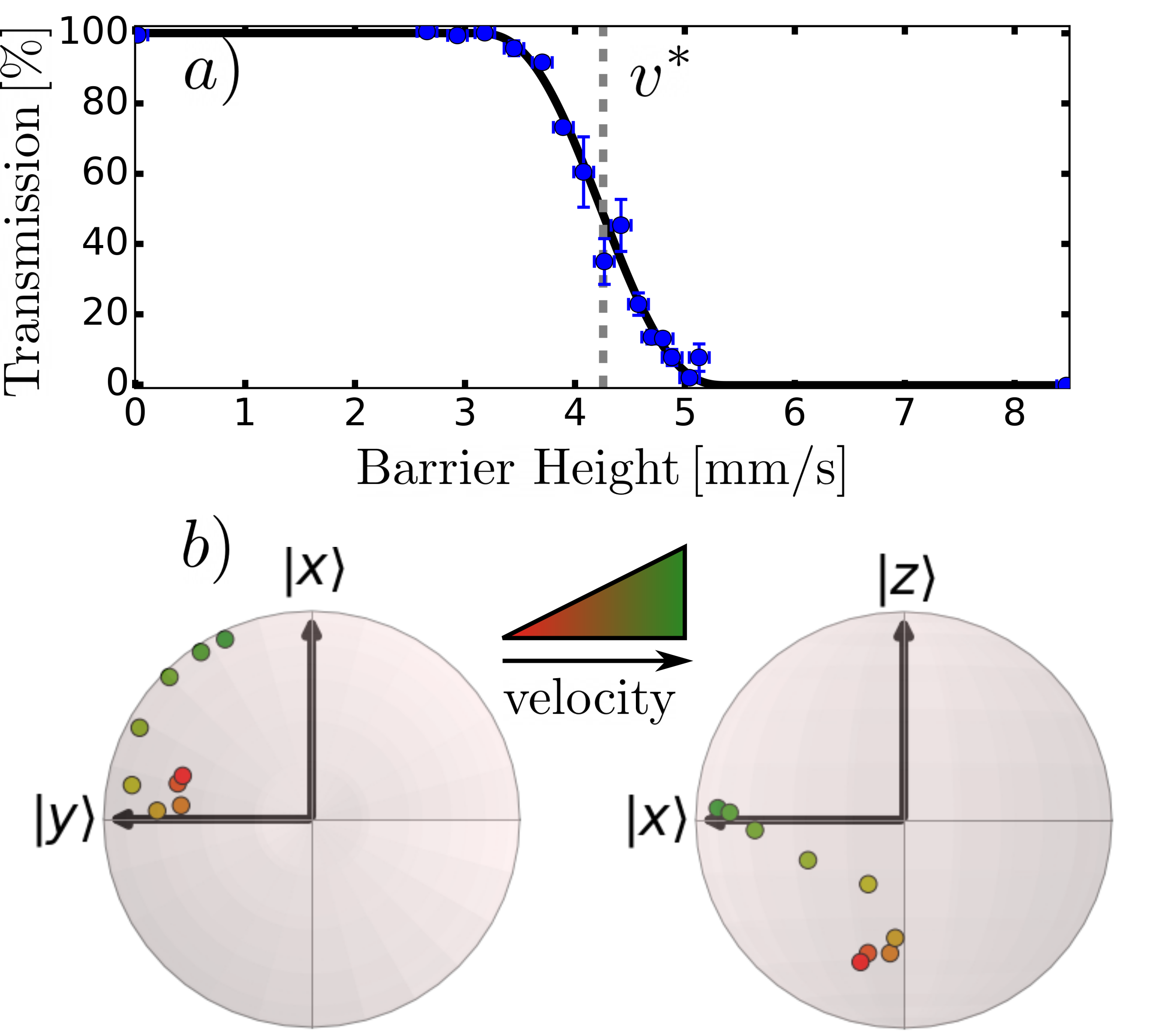}
\caption{\label{fig:knife} Typical data from an experimental run. a) Transmission of a wavepacket with average incident velocity $v^*$ through barriers of varying heights measures the width of the incident atomic cloud's velocity distribution. b) The components of the Bloch vector for the transmitted wavepacket are extracted for wavepackets with different incident velocities.}
\end{figure}

We implement a Larmor time measurement in a Bose-Einstein condensate of $^{87}$Rb atoms. The long wavelengths and slow dynamics of ultra-cold atoms make tunneling and its timescales convenient for experimental study. Figure \ref{fig:setup} illustrates the sequence of our experimental procedure (see Supplementary for further details of advances on previous methods \cite{Ramos2020}). About $3000$ atoms are initially (Fig. \ref{fig:setup}a) condensed in the $\ket{F=2,m_F=2}$ state of the $5S_{1/2}$ ground orbital of rubidium and confined in a \SI{1054}{\nano\meter} crossed optical dipole trap (ODT). An elongated dipole trap forms a quasi-1D waveguide for the scattering event. To construct a wavepacket suitably narrow in momentum such that transmission through our barrier has a minimal contribution from classical spilling, we perform matter-wave lensing on the atomic cloud (Fig. \ref{fig:setup}b) and achieve velocity widths close to \SI{0.3}{\mm\per\second}, corresponding to a thermal de Broglie wavelength of approximately \SI{5}{\micro\m}. The atoms are `kicked' by the same beam used in the initial crossed ODT. The resulting velocity spread is measured concurrently with the Larmor experiment using a technique analogous to a knife-edge calibration of an optical beam's transverse extent \cite{Ramos_knife} (Fig. \ref{fig:knife}a). This technique is also used to calibrate the barrier's height, which is the peak potential energy (or equivalent velocity) of the barrier and equal to the incident energy at which half the atoms are transmitted.

A variable-duration magnetic field gradient pushes the atomic wavepacket toward a  \SI{421.38}{\nano\meter} beam of light, which acts as a repulsive barrier for the atoms.  The barrier intersects the waveguide near the center of its longitudinal potential (Fig. \ref{fig:setup}c). In this configuration, the wavepacket has a single opportunity to tunnel through the barrier.  The mean velocity at which the cloud collides with the barrier is extracted from the motion in the absence of the barrier, accounting for the acceleration imparted by the waveguide. Along $\hat{y}$, the barrier has a Gaussian profile with $1/e^2$ radius of \SI{1.3}{\micro\m}, permitting significant tunneling, while the transverse dimensions, $d_z=\SI{50}{\micro\m}$ and $d_x=\SI{8}{\micro\m}$, ensure that the atomic wavepacket collides with an essentially uniform barrier.

\begin{figure}[t]
\includegraphics[width=\columnwidth]{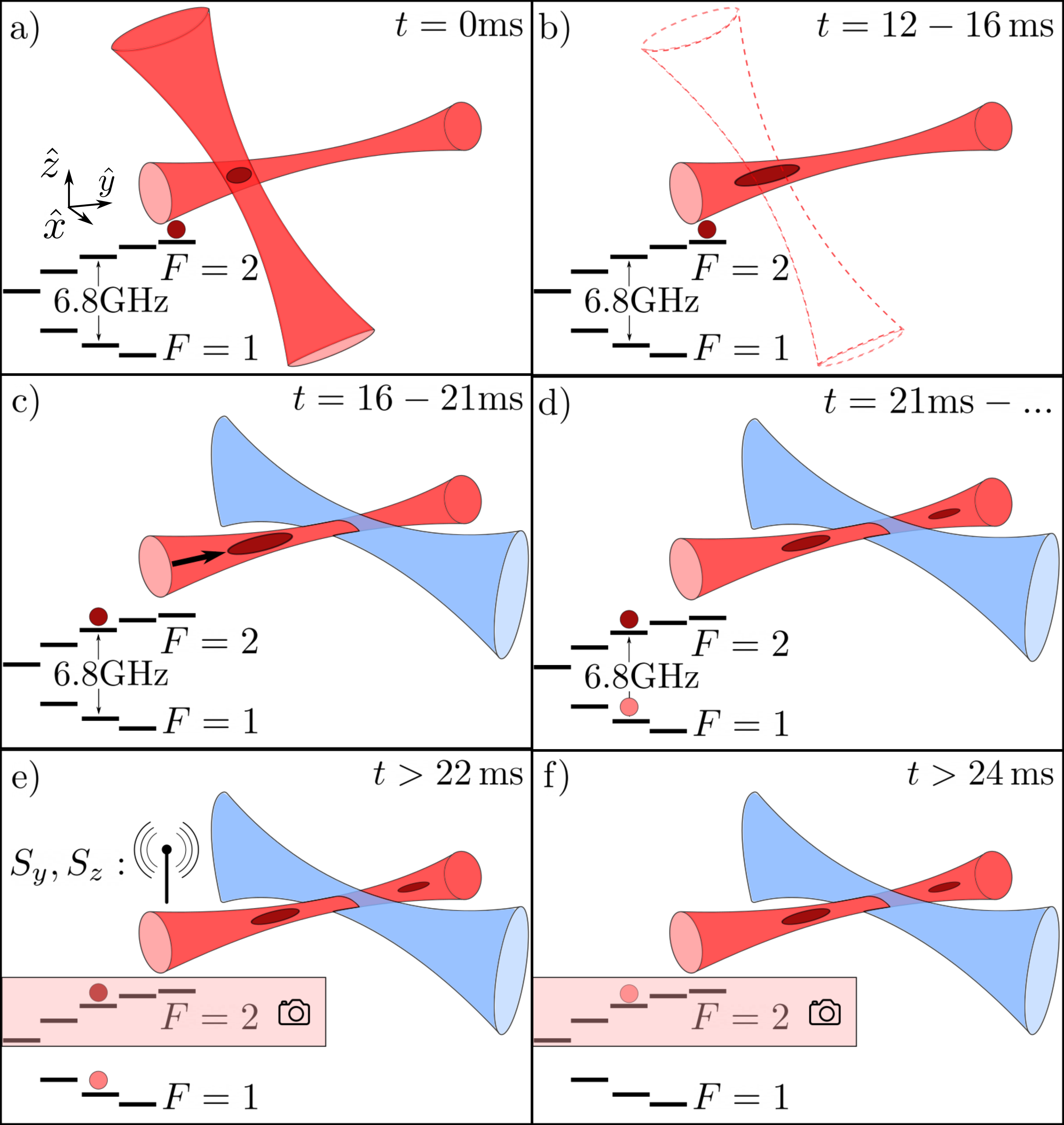}
\caption{\label{fig:setup} 
The experimental procedure: a) A cloud of $^{87}\text{Rb}$ atoms is condensed in the $\ket{F=2,m_F=2}$ state and confined at the intersection of two optical dipole traps, one of which is elongated and forms a quasi-1D system for the collision. b) Matter-wave lensing prepares an atomic wavepacket narrow in momentum. c) A variable-duration magnetic field gradient pushes the atomic wavepacket towards a \SI{421.38}{\nano\meter} optical barrier. Before reaching the barrier, the spins of the atoms are transferred to the $\ket{F=2,m_F=0}$ state via ARP.  d) A two-photon Raman transition between the clock states of the hyperfine manifold implements the Larmor measurement during the collision with the barrier. e$\&$f) Well after the collision, the $\langle S_x\rangle$,$\langle S_y\rangle$, and $\langle S_z\rangle$ components of the net magnetization vector are obtained by sequential imaging of the hyperfine populations. Measurements of the $\langle S_y\rangle$ and $\langle S_z\rangle$ components are preceded by $\pi/2$ MW rotations about the z- and y-axes of the Bloch sphere, respectively. ARP transfers the $\ket{F=1,m_F=0}$ population to $\ket{F=2,m_F=0}$ prior to the second absorption image.}
\end{figure}

Before reaching the barrier, the atoms are transferred to the $\ket{F=2,m_F=0}$ state via an adiabatic rapid passage (ARP) sweep of a radiofrequency field. The clock states of the hyperfine manifold serve as an effective spin-1/2 system for the Larmor measurement, where $\ket{F=2,m_F=0}\equiv\ket{x}=(\ket{\uparrow}+\ket{\downarrow})/\sqrt{2}$ and $\ket{F=1,m_F=0}\equiv\ket{-x}=(\ket{\uparrow}-\ket{\downarrow})/\sqrt{2}$, in the notation of Fig. \ref{fig:larmorthought}. The Larmor probe is implemented using the same beam of light as the barrier.  A \SI{6.8}{GHz} electro-optic modulator performs phase modulation on the light, and an etalon with a FWHM of \SI{12}{GHz} filters out unwanted sidebands. Together, the carrier and a single sideband couple the $\ket{\pm x}$ states via a two-photon Raman transition, acting as a pseudo-magnetic field pointing along the z-axis of the Bloch sphere.  A $\SI{1}{G}$ magnetic field pointing along $-\hat{x}$ sets the quantization axis for the atoms, and the light is circularly polarized to drive $\sigma^+ - \sigma^+$ transitions.  We use the rotation angles experienced by high-velocity clouds, which are transmitted classically over the barrier, to calibrate the effective Larmor frequency of the Raman beams.

After the wavepacket has left the barrier region, the angles of rotation depicted in Fig. \ref{fig:larmorthought} are extracted from spin-tomography of the net magnetization vector of the transmitted cloud. $\langle S_x\rangle$ is measured by counting populations in the $\ket{F=2,m_F=0}$ and $\ket{F=1,m_F=0}$ states.  To measure $\langle S_y\rangle$ and $\langle S_z\rangle$, we use a $\pi/2$ microwave (MW) pulse to rotate the axis of interest onto the axis given by the spin state populations before imaging (Fig. \ref{fig:setup}e).  The angle of the torque vector in the yz-plane about which this rotation occurs is controlled by the phase relationship between the barrier's Larmor rotation and the MW pulse. Before each data run, we calibrate the phase that aligns the torque vector with the y-axis of the Bloch sphere by `zeroing’  $\langle S_z\rangle$ for a wavepacket with velocity well above the barrier height, where transmission is not biased by spin state.


Atoms are counted in situ by sequential absorption imaging of the $\ket{F=2,m_F=0}$ and $\ket{F=1,m_F=0}$ populations with resonant light addressing only the $F=2$ manifold (Fig. \ref{fig:setup}e and f).  This keeps the atomic density high during imaging, allowing us to reliably count tens of atoms. Atoms in the $F=2$ manifold in the first image heat up due to scattering of resonant light and escape from the waveguide. A MW ARP transfers atoms from the $\ket{F=1,m_F=0}$ to the $\ket{F=2,m_F=0}$ state, immediately after which a second image is taken. This sequence can count relative atom number between the images with better than $98\%$ fidelity.

Figure \ref{fig:knife} presents data from one experimental run. During each run, we calibrate the width of the atomic cloud's velocity distribution for one choice of mean incident velocity, $\SI{4.26\pm0.06}{\mm\per\s}$, henceforth referred to as $v^*$. Figure \ref{fig:knife}a shows the transmission of a wavepacket with incident velocity $v^*$ through barriers of different heights. The width of this transmission profile is a measure of the atomic cloud's velocity spread (see Supplementary for details of fit function) \cite{Ramos_knife}.  

\begin{figure}[t]
\includegraphics[width=\columnwidth]{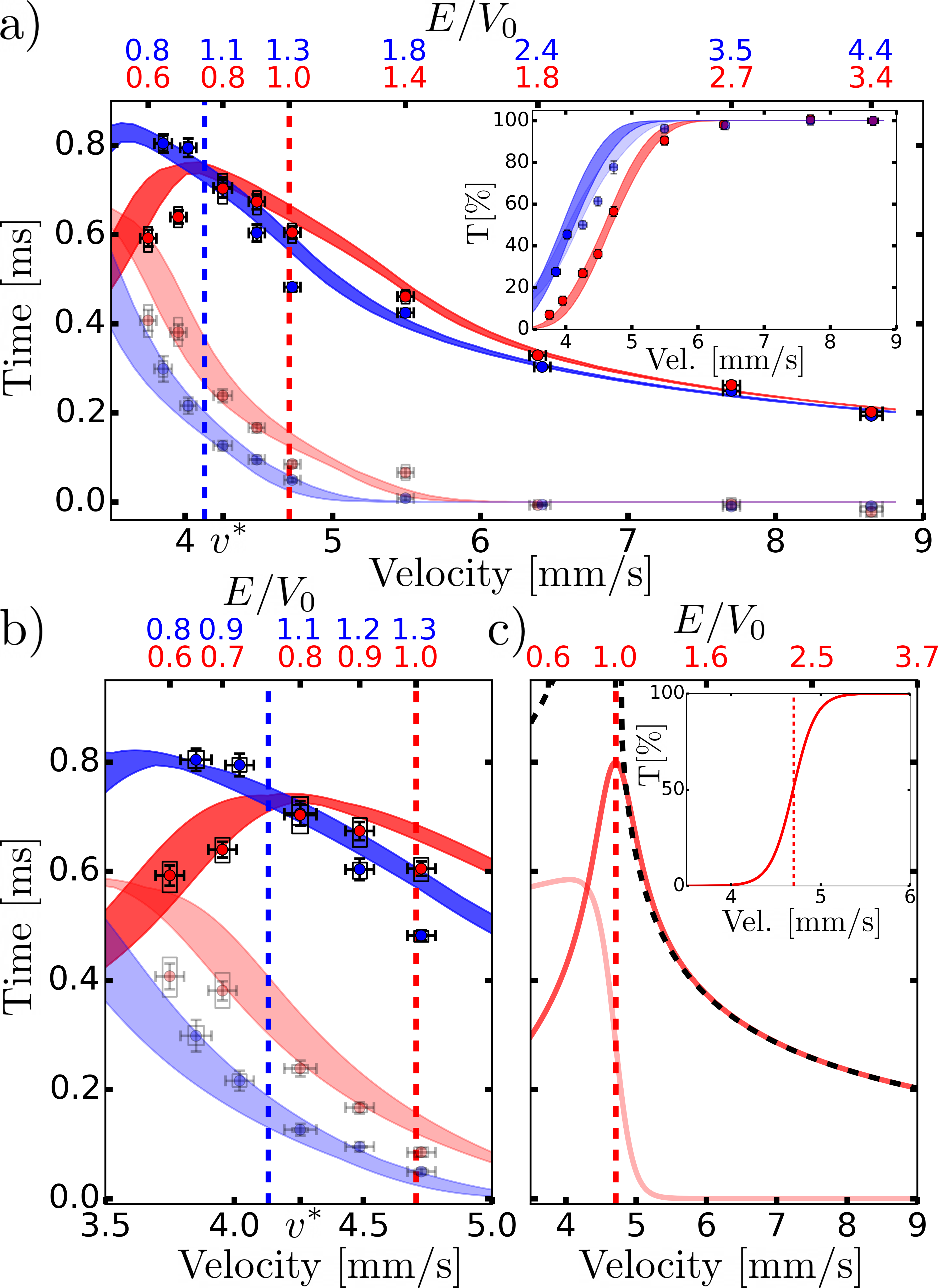}
\caption{\label{fig:times} Larmor times. a) Measured Larmor times $\tau_y$ (solid markers) and $|\tau_z|$ (faded markers) for two different barrier heights, \SI{4.71\pm0.05}{\mm\per\s} (red) and  \SI{4.13\pm0.08}{\mm\per\s} (blue), depicted by the vertical dashed lines, versus incident velocity (lower axis) and normalized energy (upper axis).  Statistical and systematic errors are given by the bars and rectangular outlines, respectively, which are sometimes smaller than the markers. Colored bands show GP simulations of the Larmor experiment, bounded by the uncertainties in the measured barrier heights and in the velocity width at $v^*$. Inset) Experimental and simulated transmission.  Some of the above-barrier data for the lower barrier height set had an atypically large Rabi frequency, \SI{808\pm12}{Hz}. While this modifies the transmission, indicated by the faded blue markers and bands, it does not significantly affect these times. b) Low energy data of (a), now compared to weak-value calculations weighted by the inferred velocity profile of the transmitted wavepacket. The shaded regions are bounded as in (a). c)  Monochromatic weak-value calculations of the Larmor times (red) for a Gaussian barrier of height \SI{4.71}{\mm\per\s} and the semiclassical expression \cite{buttiker1982traversal} used to calibrate the Larmor frequency above the barrier (see Supplementary). Inset) Transmission profile.}
\end{figure}

The 2D projections of the Bloch sphere in Fig. \ref{fig:knife}b show the rotation angles from a Larmor measurement. We perform constrained maximum-likelihood estimation of the net magnetization vector of transmitted atoms given the raw tomographic data extracted from the absorption images.  
In addition, we compensate for drifts in the relative phase between the MW pulse and the Larmor rotation, which we monitor throughout the experiment. These drifts typically require us to apply a \SI{10}{\degree} to \SI{20}{\degree} rotation to the $\langle S_y\rangle$ and $\langle S_z\rangle$ results. The in-plane angles of rotation for the two fastest incident wavepackets are used to determine the Rabi frequency for each run, which always fell in the range \SI{150}{\hertz}-\SI{350}{\hertz} (with the exception of some of the above-barrier data indicated in the inset of Fig. \ref{fig:times}a).

Figure \ref{fig:times} presents the Larmor times versus the incident velocity of wavepackets colliding with barriers of two different heights, \SI{4.71\pm0.05}{\mm\per\s} (red) and  \SI{4.13\pm0.08}{\mm\per\s} (blue), indicated by the vertical dashed lines.  The error bars on the markers represent the rms width of the data's distribution, while the rectangular outlines depict the systematic errors due to the aforementioned compensations to the spin components as well as the uncertainty in the measured Rabi frequency. Below the barrier we observe two important features of the $\tau_y$ data:
1) $\tau_y$ decreases with decreasing incident energy, a \SI{2}{\sigma} and \SI{4}{\sigma} result considering the high-barrier data, and 2) $\tau_y$ increases as the barrier height decreases, an \SI{8}{\sigma} and \SI{6}{\sigma} result considering the two lowest energies of each set. The opposite behavior is shown by $|\tau_z|$, signifying that the back-action of the measurement increases for higher barriers and lower incident velocities in the measured energy range.

We compare our data to two theoretical calculations.  The shaded regions in Fig. \ref{fig:times}a describe one-dimensional Gross-Pitaevskii (GP) simulations \cite{Bao2003} of the Larmor experiment in the presence of the longitudinal potential of the waveguide, matching the measured rms velocity width \SI{0.35\pm0.03}{\mm\per\s} at $v^*$.  The regions are bounded by the uncertainty in the measured velocity width and barrier height for each dataset.  These simulations model the changing velocity profile of the wavepacket resulting from the interplay of the repulsive atomic interactions and the confining potential of the waveguide before atoms reach the barrier (see Supplementary for details). For our experimental parameters, wavepackets with slower incident velocities develop smaller velocity widths. The GP simulations explain two key features of our results. A nonzero $\tau_z$ appears for velocities above the barrier, contrary to the predictions of the ideal monochromatic theory, due to the portion of the velocity distribution near or just below the barrier height. In addition, we see a significant decrease in $\tau_y$ and increase in $|\tau_z|$ for the slowest wavepackets, whose velocity profile is narrow enough that transmission is dominated by energies below the barrier. The GP simulations indicate that the rms velocity width of the wavepacket drops by \SI{0.06}{\mm\per\s}, comparing the width inferred from simulations of the slowest incident wavepacket and our measured value. In order to find a lower bound for the fraction of atoms that tunneled through the barrier, we use the conservative hypothesis that wavepackets slower than $v^*$ had the same velocity width as the measured value.

With this premise, we compare the low-energy data to weak-value calculations of the Larmor times, as given by Eq. \ref{eq:tauy}.  Figure \ref{fig:times}c shows monochromatic calculations with a Gaussian barrier of height \SI{4.71}{\mm\per\s}, whereas Fig. \ref{fig:times}b compares the below-barrier data to the average time calculated for the spread of velocities transmitted through both barriers. As before, the shaded regions are bounded by the uncertainties in the measured velocity width and barrier height for each data set.  We observe close agreement between the weak-value calculations and our data as well as with GP simulations. In this experiment atomic interactions evidently played little role in the transmitted times, and so the association of the data with the conditional dwell time inside the barrier is supported. In contrast, the measured Larmor times for the reflected atoms show differences from the predictions of the non-interacting weak-value theory. We believe that the mean-field energy due to atomic interactions in the high density regions in front of the barrier are important to understanding those data, which will be the topic of forthcoming work.

For the data set with a higher barrier, we estimate from the measured velocity width that at least \SI{90}{\percent} of atoms tunneled for the lowest incident energy measured and \SI{65}{\percent} of the interaction between the Larmor probe and the transmitted wavepacket occured in a forbidden region.  For the wavepacket incident at $v^*$, we deduce that \SI{50}{\percent} of transmitted atoms tunneled through the higher barrier and \SI{20}{\percent} of the interaction between Larmor probe and transmitted wavepacket occurred in a forbidden region.  The mean transmitted velocity for this data point is within uncertainty of the height of the barrier, which is consistent with the decrease in $\tau_y$.  As shown in Fig. \ref{fig:times}c, in the monochromatic limit $\tau_y$ reaches its maximum at the velocity matching the height of a Gaussian barrier and is roughly symmetric for nearby velocities about this peak.  Thus, given the symmetry of the transmission profile of a Gaussian barrier (\ref{fig:times}c inset) and a smooth incident velocity profile, $\tau_y$ is only expected to decrease for energies below the barrier once the energy is low enough that a majority of atoms in fact tunnel through the barrier.  Hence, both the measured velocity profile of the transmitted wavepackets and the decrease in $\tau_y$ are consistent with a majority of atoms having tunneled for the lowest incident energies.  This was not observed in \cite{Ramos2020} given the larger and more uncertain velocity widths there, as well as the omission of some of the systematic effects described here.

The work here shows that tunneling generally takes less time when the tunneling event has a lower probability to occur due to a larger energy deficit below the barrier. We observe the characteristic decrease of $\tau_y$ below the barrier, which for a Gaussian barrier indicates that transmission is dominated by tunneling.  Additionally, we witness the increase in $\tau_y$ for lower barriers and conclude through comparisons with theory that atomic interactions play little role in our measured times in transmission. Future work will study the distinct histories of transmitted and reflected particles, not only due to the role of atomic interactions, but more fundamentally because of the differing forbidden regions these subensembles probe.

\begin{acknowledgments}
The authors would like to thank Kehui Li for improvements to the absorption image analysis, Joseph McGowan, Nick Mantella, and Aharon Brodutch for critical reading of the manuscript, as well as Ram\'on Ramos and Isabelle Racicot for experimental developments that culminated in \cite{Ramos2020}. This work was supported by NSERC and the Fetzer Franklin Fund of the John E. Fetzer Memorial Trust. A.M.S is a Fellow of CIFAR.
\end{acknowledgments}


\bibliography{apssamp}

\pagebreak
\widetext
\begin{center}
\textbf{\large Supplementary Material: Observation of the Decrease of Larmor Tunneling Times with Lower Incident Energy}
\end{center}

\input{supplementary.tex}

\end{document}

%% file: supplementary.tex
\section{Experimental Procedure}
\subsection{Atomic Wavepacket Preparation\label{sec:wavepacket_prep}}
\begin{figure}[b]
\includegraphics[width=0.5\columnwidth]{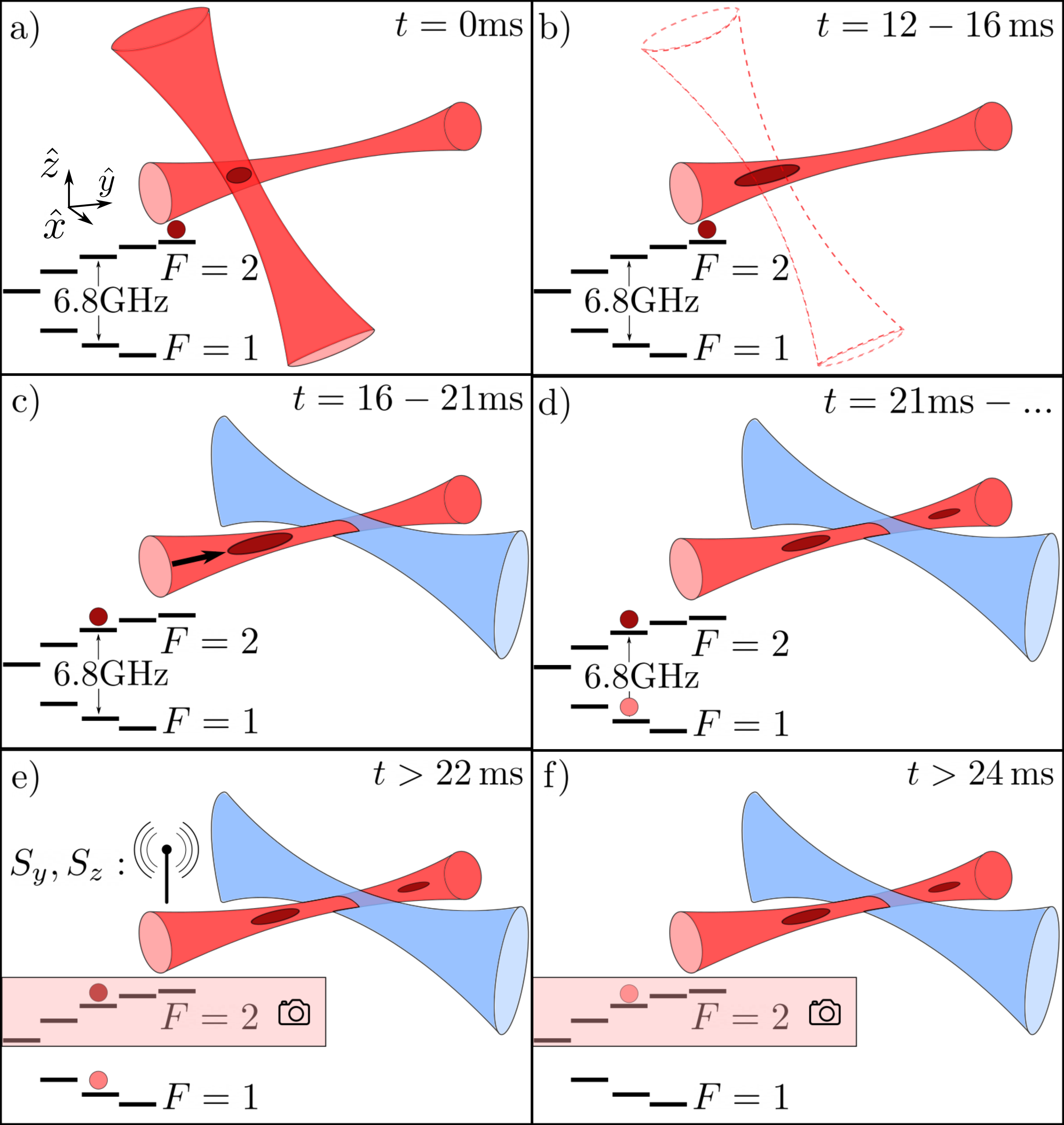}
\caption{\label{fig:setup_sup} 
The experimental procedure: a) A cloud of $^{87}\text{Rb}$ atoms is condensed in the $\ket{F=2,m_F=2}$ state and confined at the intersection of two optical dipole traps, one of which is elongated and forms a quasi-1D system for the collision. b) Matter-wave lensing prepares an atomic wavepacket narrow in momentum. c) A variable-duration magnetic field gradient pushes the atomic wavepacket towards a \SI{421.38}{\nano\meter} optical barrier. Before reaching the barrier, the spins of the atoms are transferred to the $\ket{F=2,m_F=0}$ state via ARP.  d) A two-photon Raman transition between the clock states of the hyperfine manifold implements the Larmor measurement during the collision with the barrier. e$\&$f) Well after the collision, the $\langle S_x\rangle$,$\langle S_y\rangle$, and $\langle S_z\rangle$ components of the net magnetization vector are obtained by sequential imaging of the hyperfine populations. Measurements of the $\langle S_y\rangle$ and $\langle S_z\rangle$ components are preceded by $\pi/2$ MW rotations about the z- and y- axes of the Bloch sphere, respectively. ARP transfers the $\ket{F=1,m_F=0}$ population to $\ket{F=2,m_F=0}$ prior to the second absorption image.}
\end{figure}

Figure \ref{fig:setup_sup} (reproduced from the main text) illustrates a time sequence of the experimental procedure. Initially, about $40000$ atoms are condensed in the $\ket{F=2,m_F=2}$ state of the $5S_{1/2}$ ground orbital of $^{87}\text{Rb}$ via evaporative cooling in a \SI{1054}{\nano\meter} crossed optical dipole trap (ODT).  While the atoms are already in a pure BEC, we perform additional evaporation, reducing the atom number to about $3000$ in order to limit the interaction energy of the atomic cloud and thereby limit the necessary `cooling' in later stages.  One of the dipole traps, which we refer to as the `waveguide,' is elongated and forms a quasi-1D geometry for the collision with the barrier.  The waveguide has a $1/e^2$ radius of $\sigma_{x,z}=\SI{14.5\pm0.2}{\micro\m}$, a Rayleigh range of $z_R=\SI{620\pm20}{\micro\m}$, and trap frequencies of $\nu_{x,z}=\SI{250\pm10}{\hertz}$, and $\nu_{y}=\SI{2.5\pm0.1}{\hertz}$.

We perform matter-wave lensing on the atomic cloud (Fig. \ref{fig:setup_sup}b) and achieve velocity widths of around $\SI{0.35}{\mm\per\second}$. The power in the crossing beam is ramped down over the course of \SI{12}{\ms}, after which the cloud has expanded in the waveguide to about five times its initial size and interaction energy has been almost entirely converted into kinetic energy.  The atomic cloud is `kicked' by the same crossing beam via a pulse that is ramped on and off in \SI{4}{\ms}. Importantly, each of these steps is done adiabatically with respect to the transverse trap frequencies of the waveguide in order to minimize motion transverse to the propagation direction, $\hat{y}$, and the velocity distribution remains Thomas-Fermi as seen in time-of-flight. The velocity spread of the wavepacket is measured concurrently with the Larmor experiment by scanning the power in the optical barrier, reconstructing a histogram of the atomic velocity profile (Sec. \ref{sec:app_knife}).

A variable-duration magnetic field gradient along the longitudinal direction of the waveguide sets the mean velocity for the atomic wavepacket. The duration of the magnetic push is between \SI{1}{\ms} and \SI{5}{ms} for the incident velocities studied here.  The mean velocity at which the cloud impacts the barrier is calibrated by fitting the motion of the average position of the atomic cloud as it propagates in the waveguide without the barrier. The acceleration supplied by the longitudinal potential of the waveguide is taken into account.  After the preparation stages, the wavepacket has a one-dimensional Thomas-Fermi radius of about \SI{60}{\micro\meter} and is about \SI{150}{\micro\meter} from the barrier, which intersects the waveguide near the center of its longitudinal potential.

\subsection{Barrier and Larmor Clock Implementation}
The barrier is created from a focused \SI{421.38}{\nano\meter} beam of light (Fig. \ref{fig:setup_sup}c).  This wavelength is about \SI{1.8}{\tera\hertz} red-detuned from the $6P_{3/2}$ transition, \SI{0.5}{\tera\hertz} blue-detuned from the $6P_{1/2}$ transition, and \SI{330}{\tera\hertz} blue-detuned from the $5P_{3/2}$ transition.  The combined AC stark shift from these transitions results in a repulsive interaction with the atoms. Along the direction of the waveguide, the barrier has a Gaussian profile with $1/e^2$ radius of \SI{1.3}{\micro\m} and a Rayleigh range of $\SI{8}{\micro\m}$ along its axis of propagation. In the z-direction, the beam is scanned over a $\SI{50}{\micro\m}$ range by a frequency-modulated acousto-optic deflector, with a modulating waveform with a central frequency of \SI{133}{\kilo\hertz}. The time-average of this motion `paints' a uniform potential for the atoms, ensuring that the wavepacket collides with a barrier homogeneous along $\hat{z}$.  In the 1D geometry the wavepacket has a single opportunity to tunnel through the barrier and with a few milliwatts of power we can create Gaussian barriers with peak potential energies up to $k_B\cdot$\SI{250}{\nano\kelvin}, which is a height that matches a $^{87}\text{Rb}$ atom moving at nearly \SI{7}{\mm\per\s}.

An effective spin-1/2 system is encoded in the clock states ($m_F=0$) of the two hyperfine levels of the $5S_{1/2}$ orbital, with the $\ket{F=2,m_F=0}$ state acting as the $\ket{x}$ state in the Bloch sphere picture. A radiofrequency adiabatic rapid passage (ARP) sweep is used to transfer the atoms from $\ket{F=2,m_F=2}$ to $\ket{F=2,m_F=0}$ before the atoms reach the barrier.  In the presence of a \SI{20}{G} magnetic field, the $F=2$ manifold has a significant quadratic Zeeman shift, allowing the $\ket{F=2,m_F=2}\rightarrow\ket{F=2,m_F=1}$ and $\ket{F=2,m_F=1}\rightarrow \ket{F=2,m_F=0}$ transitions to be addressed independently during the frequency sweep.  The ARP has an overall efficiency of $\SI{95}{\percent}$ and atoms not in the $m_F=0$ state are pushed away with a strong magnetic field gradient so that they do not pollute the subsequent absorption imaging.  

The clock states of the two hyperfine manifolds are addressed by a microwave (MW) oscillation at \SI{6.8}{GHz} (more precisely, the frequency found before each experimental run as described below) imprinted on the barrier beam via phase modulation from an electro-optic modulator.  The modulation depth is low so that measurable optical power is only found in the carrier and first-order sidebands.  These sidebands are out of phase and therefore would lead to destructive interference when used with the carrier to drive two-photon transitions.  An etalon with a FWHM of \SI{12}{GHz} filters out one of the first order sidebands, leaving the carrier and a single sideband with about $\SI{3}{\%}$ of the carrier's optical power.  Together, the carrier and single sideband couple the $\ket{\pm x}$ states via a two-photon Raman transition, acting as a pseudo-magnetic field pointing along the z-axis of the Bloch sphere.  The light is circularly polarized to drive $\sigma^+ - \sigma^+$ transitions. 

After the spin preparation, the magnetic field is decreased to \SI{1}{G} along $-\hat{x}$, maintaining a quantization axis for the Zeeman sublevels, parallel to the propagation direction of the Raman beams. Since the transition frequency between the clock states depends quadratically on the magnetic field, this weaker field reduces the sensitivity to noise.  Nevertheless, we find that between data runs this transition frequency shifts by a few hertz, corresponding to a \SI{1}{mG} $-$ \SI{10}{mG} drift.  If not corrected, this small detuning leads to dephasing of the atomic spins from the MW oscillation coupling the clock states, between the time they exit the Raman beams and the moment the measurement of the $S_y$ or $S_z$ components is performed.  This detuning acts as a torque vector on the Bloch sphere that would rotate $S_y$ and $S_z$ by tens of degrees, depending on the incident velocity.  Here, we perform Ramsey measurements before every experimental run to find resonance with an uncertainty of \SI{1}{\hertz} and thus reduce the possible influence of this systematic shift to less than \SI{5}{\degree} of rotation to the $S_y$ and $S_z$ components.

\subsection{Spin Tomography and Imaging\label{sec:tomography}}
On a given experimental cycle, we measure one component --- either $\langle S_x\rangle$, $\langle S_y\rangle$, or $\langle S_z\rangle$ --- of the net magnetization vector of the transmitted atomic cloud.  While $\langle S_x\rangle$ is measured simply by counting populations in the $\ket{F=2,m_F=0}$ and $\ket{F=1,m_F=0}$ states, a $\pi/2$ rotation about the z- and y-axes of the Bloch sphere must be performed before measuring the hyperfine populations in order to measure the $\langle S_y\rangle$ and $\langle S_z\rangle$ components, respectively.  This rotation is performed after the collision with the barrier by a MW pulse with Rabi frequency of \SI{727.0\pm0.8}{\hertz} (Fig. \ref{fig:setup_sup}e).  The angle of the torque vector in the yz-plane is controlled by the phase relationship between the MW oscillation imprinted on the barrier beam and the subsequent MW pulse.  We calibrate this phase difference before each data run by measuring the out-of-plane spin component as a function of the relative phase of the MW pulse, for a wavepacket with velocity well above the barrier height.  Since high-velocity wavepackets pass over the barrier regardless of their spin state, the phases for which the out-of-plane spin component is zero identify when the torque vector lies along the y-axis of the Bloch sphere (i.e. an $\langle S_z\rangle$ measurement).  

The spin tomography is done in situ by sequentially imaging the $\ket{F=2,m_F=0}$ and $\ket{F=1,m_F=0}$ populations.  We perform resonant absorption imaging on the $\ket{F=2,m_F=0}$ population for \SI{100}{\micro\s}, after which the atoms have gained a velocity kick sufficient to eject them from the waveguide (Fig. \ref{fig:setup_sup}e).  Atoms heated in this first absorption image cannot be seen in subsequent images more than \SI{1.5}{\ms} later.  After the first image, a \SI{3}{\ms} MW ARP transfers atoms from the $\ket{F=1,m_F=0}$ to the $\ket{F=2,m_F=0}$ state, immediately after which a second absorption image is taken.  Finally, \SI{6}{\ms} after the second absorption image (limited by the transfer rate of our camera), a reference image is taken.  We have verified that this sequence can count relative atom number between the images with better than $98\%$ fidelity.

\section{Data Analysis}
\subsection{Velocity Width Measurements\label{sec:app_knife}}
For a given mean incident velocity, $v_0$, we measure the velocity width of the atomic wavepacket via a scan of the barrier height. We start with a 1D velocity distribution, obtained by integrating a 3D momentum-space Thomas-Fermi distribution over the two transverse dimensions, and we fit the transmission profile to the integral of this distribution: 
\begin{equation}
    1 - A\int_0^{v_b} dv_b' \begin{cases}    \big(1-\frac{(v_b'-v_0)^2}{v_R^2}\big)^2 & \text{for $|v'_b - v_0|<v_R$} \\
                                   0 & \text{otherwise,}
  \end{cases}
\end{equation}
where $A$ normalizes the 1D velocity distribution, $v_b$ is the velocity matching the energy of the barrier's peak, and $v_R$ is the Thomas-Fermi radius of the velocity profile.  While $A$ and $v_R$ are fit parameters, $v_0$ is known from the  wavepacket velocity calibrations described in Sec. \ref{sec:wavepacket_prep}. For a Gaussian barrier, half of the atoms are transmitted when the incident energy matches the barrier's peak, which allows us to calibrate $v_b$ for a given barrier beam intensity, measured by a photodiode, using transmission profiles for a few different incident velocities. Transmission due to tunneling through a Gaussian barrier of our measured spatial width contributes an rms width of \SI{0.21}{\mm\per\s} to the transmission profile.  To a good approximation, the measured velocity width is the quadrature sum of the width due to tunneling through the barrier and the rms velocity width of the atomic cloud \cite{Ramos_knife}. 

Two competing influences cause the velocity width of the atomic wavepacket to change with time. Remnant interaction energy of the atomic cloud after the matter-wave lensing stage causes the velocity width to grow. In addition, the harmonic longitudinal potential of the waveguide causes the phase-space distribution of the cloud to rotate.  The effect of the waveguide is stronger, and in the time atoms spend prior to collision with the barrier, for all incident velocities used in the experiment, the velocity width of the cloud decreases.  As a result, wavepackets with slower incident velocities develop smaller velocity widths before they collide with the barrier.

In order to ensure that transmission at low velocities has a large contribution from tunneling and is not overwhelmed by classical spilling, we measure the velocity width of the wavepacket for one of the slower velocities, \SI{4.26\pm0.06}{\mm\per\second}, referred to as $v^*$ in the main text, concurrently with the Larmor experiment.  Given the trap frequencies of the waveguide, the initial distance from the barrier, and the atom number, we can simulate the changing velocity profile of the atoms via Gross-Pitaevskii simulations. The simulations show that our fastest and slowest wavepackets had velocity profiles with rms widths of about \SI{0.57}{\mm\per\s} and \SI{0.29}{\mm\per\s}, respectively.

\subsection{Spin Measurements}
The tomography described in Sec. \ref{sec:tomography} measures $\langle S_x\rangle$, $\langle S_y\rangle$, and $\langle S_z\rangle$ of the net magnetization vector of the transmitted cloud. When the magnetization vector happens to be situated close to one of the measured axes, experimental uncertainty can cause some of the data to be outside of the Bloch sphere, particularly when atom number is low in one of the absorption images. Given the knowledge that physically real Bloch vectors exist on or inside the Bloch sphere, we update our estimates of the Bloch vectors by constraining the data's distribution to lie inside the sphere (i.e. constrained maximum-likelihood estimation).

Furthermore, we compensate for miscalibrations of the $S_y$ and $S_z$ measurement axes.  During a data run, the etalon used for filtering the Raman beams experiences temperature drifts, which affect the relative phase between the Larmor rotation and the MW source. These drifts lead to errors in the orientation of the torque vectors we apply when we attempt to project $\langle S_y\rangle$ and $\langle S_z\rangle$ onto the axis given by the hyperfine populations. We monitor this drift via an $\langle S_z\rangle$ measurement of a high-velocity data point as described earlier.  The corrections required are typically about a \SI{5}{\degree} rotation to the measurement axes for the $S_y$ and $S_z$ components. Additionally, phase shifts also on the order of \SI{5}{\degree} to \SI{10}{\degree} arise from changing the output amplitude of the MW source between initial calibrations and the Larmor measurement.  The uncertainties in these compensations contribute to the systematic errors reported in Fig. 4 of the main text.

\subsection{Larmor Times Calculation\label{sec:angles}}
As described in the main text, the Larmor times for tunneling particles in the limit of a weak magnetic field are given by
\begin{equation}
\label{eq:BLtaus}
\tau_y+i\tau_z=-\hbar\frac{\mathlarger{\partial}\phi}{\mathlarger{\partial} V_0}+i\hbar\frac{\mathlarger{\partial} \ln(|t|)}{\mathlarger{\partial} V_0}=i\hbar\frac{\mathlarger{\partial} \ln t}{\mathlarger{\partial} V_0},
\end{equation}
where $\phi$ is the phase of the transmission amplitude $t=|t|e^{i\phi}$, and $V_0$ is the energy of the barrier. $\tau_y$ is determined by the phase shift of the transmitted particles because the in-plane rotation angle $\theta_y$ results from \emph{precession} about the magnetic field of the Larmor measurement (i.e. relative phase accumulation of the $\ket{\uparrow}$ and $\ket{\downarrow}$ states).  On the other hand, the out-of-plane angle $\theta_z$ derives from the preferential transmission of the spin component with greater energy in the presence of the magnetic field, and so is connected to changes in the magnitude of the transmission amplitude. Since $\langle S_z\rangle$ is the normalized occupation difference between the $\ket{\uparrow}$ and $\ket{\downarrow}$ states, it is a direct measure of this biased transmission.  Accordingly, we characterize the experimental Larmor rotations following
\begin{align}
    \theta_y&=\arctan(\langle S_y\rangle / \langle S_x \rangle)\\
    \alpha_z&=\arctanh(\langle S_z\rangle)\label{eq:arctanh}.
\end{align}
In the limit of infinitesimal perturbations $\alpha_z\rightarrow\theta_z$ (i.e. the geometric out-of-plane angle), but for finite Larmor frequencies only $\alpha_z$ remains a measure of $\ln(|t_\uparrow|)-\ln(|t_\downarrow|)$, where $|t_i|$ is the magnitude of the transmission amplitude for spin state $i$.  This is clear when one considers the limit of large magnetic fields, where the out-of-plane angle saturates. On the other hand, $\theta_y$ is a measure of the phase difference $\phi_\uparrow - \phi_\downarrow$ regardless of Larmor frequency.

We use the rotation angles experienced by high-velocity atomic clouds to calibrate the effective Larmor frequency of the Raman beams.  For energies well above the barrier, a semi-classical approach is a good approximation for the time spent traversing the barrier region.  Including the spatial profile of the barrier and therefore the Larmor probe, we use the following equation to define the effective Larmor frequency $\Omega$:
\begin{equation}
\label{eq:rabi}
\theta=\Omega\int_{-\infty}^{\infty} G(y)/\sqrt{v^2-v_b^2G(y)}dy,
\end{equation}
where $\theta$ is the rotation angle experienced by a high-velocity wavepacket, $G(y)=e^{-2y^2/\sigma^2}$ is the Gaussian profile of the barrier along the waveguide direction, $v$ is the velocity of the cloud, and $v_b$ is the velocity matching the height of the barrier.  Due to the spatial dependence of the Larmor probe, our measurement is a weighted sum of the times spent in different regions of the barrier. The above definition introduces the convention that, in the limit of a vanishing barrier height, our measurement corresponds to the time spent in a region of width $\sqrt{\pi/2}\sigma$.  The uncertainty in $\Omega$ adds to the systematic errors of Fig. 4 of the main text.

\section{Theory}
We compare the data to two theoretical calculations. While time-dependent Gross-Pitaevskii (GP) simulations of the Larmor experiment provide insight into the role of atomic interactions as well as other details of the experimental implementation, time-independent calculations of the conditional dwell time in the barrier provide a comparison to weak measurement theory.
\subsection{Gross-Pitaevskii Simulations}
We perform GP simulations of the Larmor experiment using the following set of coupled equations: 
\begin{align}
\label{eq:gp1}
i\hbar\frac{\partial\psi_1}{\partial t}=\bigg[\frac{-\hbar^2}{2m}\nabla^2+V+U_{11}|\psi_1|^2+U_{12}|\psi_2|^2+\hbar\delta\bigg]\psi_1+\hbar\Omega\psi_2 \\
\label{eq:gp2}
i\hbar\frac{\partial\psi_2}{\partial t}=\bigg[\frac{-\hbar^2}{2m}\nabla^2+V+U_{22}|\psi_2|^2+U_{21}|\psi_1|^2\bigg]\psi_2+\hbar\Omega\psi_1,
\end{align}
where $V$ is the trapping and barrier potential, $\hbar\delta$ is an energy offset, $\Omega$ is the Rabi frequency of the Raman beams, and $U_{ij}=4\pi\hbar^2a_{ij}/m$ is the spin-dependent interaction parameter, with s-wave scattering length $a_{ij}$ between the $i$ and $j$ components.  Given the aspect ratio of the waveguide trap, with $\omega_y<<\omega_x,\omega_z$, the dynamics of the atomic cloud are well-described by Eqs. \ref{eq:gp1} and \ref{eq:gp2} in reduced, one-dimensional form \cite{Bao2003}, with $V(y)$ and $\Omega(y)$ having Gaussian profiles.  We use the time-splitting spectral method to solve the coupled one-dimensional GP equations.  First, we use imaginary time propagation to find the ground state of the initial trapping configuration, Fig. \ref{fig:setup_sup}a, with an energy offset to initialize the wavepacket in spin state 1, which we associate with $\ket{x}$. We then release the atom cloud from the initial crossed trap and simulate matter-wave lensing similar to that done in the experiment. We adjust the duration of the lensing stage so that the simulated velocity width agrees with the measured value.  With this simulated wavepacket preparation, we model the Larmor experiment by extracting the net magnetization vector for the transmitted cloud after the collision with the barrier in the same waveguide geometry.  The incident velocity of the wavepacket is set by a phase gradient written across the initial wavepacket after the matter-wave lensing stage.  We emulate the Raman coupling of the experiment by giving $\Omega(y)$ the same Gaussian spatial profile as the barrier and define the effective Rabi frequency as in Sec. \ref{sec:angles}.  We find that the primary effect of interactions and the waveguide potential on the Larmor times for the transmitted wavepacket is to modify the velocity profile of the incident wavepacket prior to collision with the barrier.  Given this, we see close agreement between GP simulations and the experimental measurements.

\subsection{Weak Value Calculations}
We calculate the Larmor times for transmitted particles in the limit of a weak measurement \cite{buttiker1983larmor,Ae1995_time,Ae1995_cond}.  Following the weak measurement formalism, the dwell time in the region $y$ to $y+dy$, conditioned on transmission, is given by the weak value of the projector $\Theta_{d}\equiv dy\ket{y}\bra{y}$,

\begin{align}
\label{eq:wv}
\frac{1}{j_{in}}\langle\Theta_{d}\rangle_{weak}\equiv\frac{1}{j_{in}}\frac{\bra{t}\Theta_d\ket{i}}{\braket{t|i}},
\end{align}
where $j_{in}$ is the incident flux, $\ket{i}$ is the initial incident wave, and $\ket{t}$ is the transmitted state \cite{Ae1995_time}.  The real and imaginary components of this complex quantity are equal to the Larmor times $\tau_y$ and $\tau_z$, as given by Eq. \ref{eq:BLtaus} for a magnetic field localized to the region between $y$ and $y+dy$. The equality between the Larmor times and the weak value of the projector onto the region of interest is the rationale for the interpretation of $\tau_y$ and $\tau_z$ described in the main text.

In order to find the time spent across the full extent of the barrier, one must integrate Eq. \ref{eq:wv}.  Experimentally this is done by having a Larmor probe everywhere inside the barrier region, with a Larmor frequency whose spatial profile matches the barrier's.  Accordingly, we could calculate $\int_{-\infty}^\infty dyG(y)\langle \Theta_{d}\rangle_{weak}/j_{in}$, where $G(y)$ is the Gaussian profile weighting the time spent in each region, to compare with our measurements.  Equivalently, we can calculate the derivative of $\ln(t)$ with respect to the amplitude of the entire barrier, as in Eq. \ref{eq:BLtaus}. From this perspective, it becomes clear that the Raman beams in the experiment implement a perturbation to the barrier, everywhere proportional to the barrier potential.

We use the transfer-matrix method to solve the one-dimensional, time-independent Schr\"odinger equation for a monochromatic wave, incident on a Gaussian barrier matching our experimental geometry.  From the complex transmission amplitude, $t$, we calculate the Larmor times from Eq. \ref{eq:BLtaus} as described above.  In Fig. 4b of the main text, we compare the below-barrier data to the average time, calculated in this way, for the spread of velocities transmitted through each barrier, under the assumption that the incident velocity spread equals the measured value.  As discussed in the main text, we see good agreement between the experimental data and these weak-value calculations of the Larmor times.



%